\documentclass[12pt]{article}

\usepackage{setspace}
\usepackage[margin=1in]{geometry}

\usepackage{fullpage}
\usepackage{multirow}
\usepackage{amsmath, amsfonts, amssymb, amsthm}
\usepackage{bm}
\usepackage{tikz}
\usepackage{algorithm, algorithmic}
\usepackage{mdwlist}
\usepackage{enumerate}
\usepackage[round]{natbib}
\usepackage{hyperref}
\hypersetup{colorlinks,
citecolor=blue,
linkcolor=blue,
urlcolor=black
}

\def\argmin{\mathop{\rm arg\, min}}

\newcommand{\bel}{\begin{eqnarray}\label}
\newcommand{\eel}{\end{eqnarray}}
\newcommand{\bes}{\begin{eqnarray*}}
\newcommand{\ees}{\end{eqnarray*}}
\newcommand{\bei}{\begin{itemize}}
\newcommand{\beiftnt}{\begin{itemize}\footnotesize}
\newcommand{\eei}{\end{itemize}}

\def\benu{\begin{enumerate}}
\def\eenu{\end{enumerate}}

\def\argmin{\mathop{\rm arg\, min}}
\def\real{{\mathbb{R}}}
\def\R{{\real}}

\def\E{{\mathbb{E}}}
\def\P{{\mathbb{P}}}

\def\complex{\mathop{{\rm I}\kern-.58em\hbox{\rm C}}\nolimits}
\def\pa{\partial}

\def\Var{\hbox{\rm Var}}

\def\mathbold{\boldsymbol} 

\def\ahat{\widehat{a}}

\def\bA{\mathbold{A}}

\def\bb{\mathbold{b}}\def\bhat{\widehat{b}}
\def\hbb{{\widehat{\bb}}}

\def\calD{{\cal D}}

\def\calI{{\cal I}}

\def\calP{{\cal P}}

\def\calR{{\cal R}}

\def\bV{\mathbold{V}}
\def\Vbar{{\overline V}}

\def\bX{\mathbold{X}}
\def\Xbar{{\overline X}}

\def\bbeta{\mathbold{\beta}}

\def\gammahat{\widehat{\gamma}}

\def\btheta{\mathbold{\theta}}\def\thetahat{\widehat{\theta}}\def\htheta{\widehat{\theta}}
\def\thetabar{{\overline{\theta}}}
\def\hbtheta{{\widehat{\btheta}}}

\def\muhat{\widehat{\mu}}\def\hmu{\widehat{\mu}}

\def\hsigma{\widehat{\sigma}}

\def\0{\mathbold{0}}

\allowdisplaybreaks
\newcommand{\ignore}[1]{}

\newtheorem{theorem}{Theorem}
\newtheorem{lemma}{Lemma}

\newtheorem{definition}{Definition}

\theoremstyle{definition}

\newcommand{\mytop}{{\small T}}
\newcommand\floor[1]{\lfloor#1\rfloor}

\begin{document}

\begin{center}
\Large Group-Linear Empirical Bayes Estimates for a Heteroscedastic Normal Mean
\end{center}

\newcommand\blfootnote[1]{%
  \begingroup
  \renewcommand\thefootnote{}\footnote{#1}%
  \addtocounter{footnote}{-1}%
  \endgroup
}

\begin{center}
Asaf Weinstein$^{*}$ \ \ \ Zhuang Ma$^{*}$ \ \ \ Lawrence D. Brown$^{*}$ \ \ \ Cun-Hui Zhang$^{*}$
\end{center}

\blfootnote{*Asaf Weinstein is at Stanford University, Stanford, CA 94305 (E- mail: asafw@stanford.edu). 
Zhuang Ma is at The Wharton School, University of Pennsylvania, Philadelphia, PA 19104 (E- mail: zhuangma@wharton.upenn.edu). 
Lawrence D. Brown is Miers Busch Professor and Professor of Statistics, The Wharton School, University of Pennsylvania, Philadelphia, PA 19104 (E- mail: lbrown@wharton.upenn.edu). 
Cun-Hui Zhang is Professor, Rutgers University, Piscataway NJ 08854 (E -mail:cunhui@stat.rutgers.edu)
}



\vspace{-.5in}

\begin{abstract}
The problem of estimating the mean of a normal vector with known but unequal variances introduces substantial difficulties that impair the adequacy of traditional empirical Bayes estimators. 
By taking a different approach, that treats the known variances as part of the random observations, we restore symmetry and thus the effectiveness of such methods. 
We suggest a group-linear empirical Bayes estimator, which collects observations with similar variances and applies a spherically symmetric estimator to each group separately. 
The proposed estimator is motivated by a new oracle rule which is stronger than the best linear rule, and thus provides a more ambitious benchmark than that considered in previous literature. 
Our estimator asymptotically achieves the new oracle risk (under appropriate conditions) and at the same time is minimax. 
The group-linear estimator is particularly advantageous in situations where the true means and observed variances are empirically dependent. 
To demonstrate the merits of the proposed methods in real applications, we analyze the baseball data used in Brown (2008), where the group-linear methods achieved the prediction error of the best nonparametric estimates that have been applied to the dataset, and significantly lower error than other parametric and semi-parametric empirical Bayes estimators. 

\end{abstract}


\newpage

\section{Introduction} \label{sec:introduction}

\noindent Let $\bX = (X_1,...,X_n)^\mytop$, $\btheta = (\theta_1,..., \theta_n)^\mytop$ and $\bV = (V_1, ..., V_n)^\mytop$, and suppose that
\bel{eq:hetero-normal}
X_i |(\theta_i , V_i) \sim N(\theta_i,V_i)
\eel
independently for $1\leq i\leq n$, where $\btheta$ and $\bV$ are deterministic. 
In the heteroscedastic normal mean problem, the goal is to estimate the vector $\btheta$ based on $\bX$ and $\bV$ under the (normalized) sum-of-squares loss
\bel{eq:loss-quadratic}
L_n(\btheta,\hbtheta) = n^{-1}\| \hbtheta - \btheta \|^2 = n^{-1} \sum_{i=1}^n (\thetahat_i - \theta_i)^2. 
\eel
Hence we assume that in addition to the random observations $X_1,...,X_n$, the variances $V_1,...,V_n$ are available. 
Allowing the values of $V_i$ to be different from each other extends the applicability of the homoscedastic Gaussian mean problem to many realistic situations.  
A simple but common example is the design corresponding to a one-way homoscedastic Analysis of Variance with unequal cell counts; 
here $X_i$ represents the mean of the $n_i$ i.i.d. $N(\theta_i, \sigma^2)$ observations for the $i$-th sub-population, hence $V_i = \sigma^2/n_i$. 
More generally, if $Y\sim N_p(\bA\bbeta, \sigma^2I)$ with a known design matrix $\bA$, then estimating $\bbeta$ under sum-of-squares loss is equivalent to estimating $\btheta$ in \eqref{eq:hetero-normal} where $n=rank(\bA)$ and $X_i$ and $V_i/\sigma^2$ are determined by $\bA$ \citep[see, e.g.,][section 2.9]{johnstone2011gaussian}. 
In both cases $V_i$ are typically known only up to a proportionality constant which can be substituted by a consistent estimator. 

The 
normal mean problem has been studied extensively for both the special case of equal variances, $V_i \equiv \sigma^2$, and the more general case above. 
Alternative estimators to the usual minimax estimator $\hbtheta = \bX$ have been suggested that perform better, for fixed $n$ or only asymptotically (under some conditions), in terms of the risk $R_n(\btheta, \hbtheta) = \E_{\btheta}[L_n(\btheta,\hbtheta)]$, regardless of $\btheta$. 
Here and elsewhere we suppress in notation the dependence of the risk function on $\bV$.

In the heteroscedastic case there is no agreement 
between minimax estimators and existing empirical Bayes estimators regarding how the components of $\bX$ should be shrunk relative to their individual variances. 
Existing parametric empirical Bayes estimators, which usually start by putting an i.i.d. normal prior on the elements of $\btheta$ and therefore shrink $X_i$ in proportion to $V_i$, are in general not minimax. 
And vice versa, minimax estimators do not provide substantial reduction in the Bayes risk under such priors, essentially under-shrinking the components with larger variances, and in some constructions \citep[e.g.][]{berger1976admissible} even shrink $X_i$ inversely in proportion to $V_i$. 
Nontrivial spherically symmetric shrinkage estimators that have been suggested, that is, estimators that shrink all components by the same factor regardless of $V_i$, are minimax only when the $V_i$ satisfy certain conditions that restrict how much they can be spread out. 
See \citet{tan2015improved} for a concise review of some existing estimators and references therein for related literature. 
Before proceeding, we remark that it is tempting to scale $X_i$ by $1/\sqrt{V_i}$ in order to make all variances equal; however, after applying this non-orthogonal transformation the loss needs to be changed accordingly (to a weighted loss) in order to maintain equivalence between the problems. 

There have been attempts to moderate the respective disadvantages of estimators resulting from either of the two approaches mentioned above. 
For example, \citet{xie2012sure} consider the family of Bayes estimators arising from the usual hierarchical model
\bel{eq:normal-normal}
\theta_i \stackrel{\text{iid}}{\sim} N(\mu, \gamma) \ \ \ \ \ \ \ \ X_i|\theta_i \stackrel{\text{ind}}{\sim} N(\theta_i, V_i) \ \ \ \ \ \ \ \ \ \ \ 1\leq i \leq n
\eel
and indexed by $\mu$ and $\gamma$. 
They suggest to plug into the Bayes rule,
\bel{eq:bayes-normal-normal}
\thetahat_i^{\mu, \gamma} = \E_{\mu, \gamma}(\theta_i | X_i) = X_i - \frac{V_i}{V_i + \gamma}(X_i - \mu),
\eel
values $(\muhat, \gammahat) = \argmin_{\mu, \gamma} \calR(\mu, \gamma; \bX )$ where $\calR(\mu, \gamma; \bX)$ is an unbiased estimator of the risk of $\thetahat^{\mu, \gamma}$. This reduces the sensitivity of the estimator to how appropriate model \eqref{eq:normal-normal} is, as compared to the usual empirical Bayes estimators, that use Maximum Likelihood or Method-of-Moments estimates of $\mu, \gamma$ under \eqref{eq:normal-normal}. 
On the other hand, \citet{berger1982selecting} suggested a modification of his own minimax estimator \citep{berger1976admissible}, 
that improves Bayesian performance while retaining minimaxity. 
\citet{tan2015improved} recently suggested a minimax estimator with similar properties that has a simpler form. 

While empirical Bayes estimators based on \eqref{eq:normal-normal} can be constructed so they asymptotically dominate the usual estimator \citep{xie2012sure}, 
the {\it modeling} of $\theta_i$ as identically distributed random variables is often not as well motivated 
in the heteroscedastic case as it is in the equal variances case. 
The assumption that $\theta_i$ are i.i.d. reflects, as commented by \citet{efron1973stein}, a ``Bayesian statement of belief that the $\theta_i$ are of comparable magnitude". 
But this assumption is not always appropriate. 
There are many examples where an association between the $V_i$ and the $\theta_i$ is expected: in Section \ref{sec:example} we consider batting records for Major League baseball players, where better performing players tend to also have larger numbers of at-bats (affecting the sampling variances of the observations). 
In situations where the true means and the $V_i$ are associated, modeling the $\theta_i$ as i.i.d. is not adequate. 
%
%
%
Nevertheless, symmetry can be restored in the heteroscedastic case by treating the {\it pair} $(X_i,V_i)$ as the random data. 
%
%
This observation leads us to 
develop a block-linear empirical Bayes estimator that groups together observations with similar variances and applies a spherically symmetric minimax estimator to each group separately. 

The rest of the paper is organized as follows. 
Section~\ref{sec:cd-hetero} presents the estimation of a heteroscedastic mean as a compound decision problem. 
This motivates the construction of a group-linear empirical Bayes estimator in Section \ref{sec:gl}; 
we discuss the properties of the proposed estimator and prove two oracle inequalities, which establish a sense of asymptotic optimality with respect to the class of estimators that are  ``conditionally" linear. 
Simulation results are reported in Section~\ref{sec:simulation}. 
In Section~\ref{sec:example} we apply our estimator to the baseball data of \cite{brown2008season} and compare it to some of the best-performing estimators that have been tested on this dataset. 
Proofs appear in the appendix. 


\section{A Compound Decision Problem for the Heteroscedastic Case} \label{sec:cd-hetero}
Let $\bX, \btheta$ and $\bV$ be as in \eqref{eq:hetero-normal}. 
It is convenient to think of $\btheta$ and $\bV$ as nonrandom, although the derivations below hold also when $\btheta$ or $\bV$ (or both) are random. 
In the sequel we refer to a rule $\hbtheta$ as {\it separable} if $\thetahat_i(\bX,\bV) = t(X_i,V_i)$ for some function $t: \R \times \R_+ \to \R$. 
Denote by $\calD_S$ the set of all separable rules. 
If $\hbtheta\in \calD_S$ with $\thetahat_i(\bX,\bV) = t(X_i,V_i)$, then
\bel{eq:cd-hetero}
R_n(\btheta, \hbtheta) = \frac{1}{n} \sum_{i=1}^n  \E_{\theta_i}[t(X_i, V_i) - \theta_i ]^2 = \E[t(Y,A) - \xi]^2
\eel
where the expectation in the last term is taken over the random vector $(Y,\xi,A,I)^\mytop$ distributed according to
\bel{eq:joint-dist}
\P(I = i) = 1/n,  \ \ \ \ (Y, \xi, A)|(I=i) \sim (X_i, \theta_i, V_i) \ \ \ \ \ \ \ \ \ 1\leq i \leq n.
\eel
Above, the symbol $``\sim"$ stands for ``equal in distribution". 
In words, \eqref{eq:joint-dist} says that $(\xi, A)$ have the empirical joint distribution of the pairs $(\theta_i,V_i)$; and $Y|(\xi, A) \sim N(\xi,A)$. 
Throughout the paper, when we refer to the random triple $(Y,\xi,A)$, its relation to $(X_i, \theta_i,V_i),\ 1\leq i\leq n$, is given by \eqref{eq:joint-dist}. 
The identity \eqref{eq:cd-hetero} -- a computation \`a la Robbins -- is easily verified by calculating the expectation on the right hand side by first conditioning on $I$. 
It says that for a separable estimator, the risk is  equivalent to the Bayes risk in a one-dimensional estimation problem. 
%


Now consider $\hbtheta \in \calD_S$ with $t$ linear (affine, in point of fact, but with a slight abuse of terminology we use the former term for convenience) in its first argument, that is,
\bel{eq:lin-given-v}
\thetahat_i^{a,b}(\bX, \bV) = X_i - b(V_i)[ X_i - a(V_i) ] \ \ \ \ \ \ \ \ \ 1\leq i \leq n
\eel
for some functions $a,b$. 
The corresponding Bayes risk in the last expression of \eqref{eq:cd-hetero} is 
\bel{eq:bayes-risk-ab}
r_n(a, b) := \E \Big\{Y - b(A)[Y - a(A)] - \xi\Big\}^2. 
\eel
Since 
\bel{eq:hetero-normal-cond}
Y|(\xi, A) \sim N(\xi,A),
\eel 
the minimizers of 
\bel{eq:bayes-risk-ab-cond}
r_n(a, b|v) :=  \E \Big\{ \Big( Y - b(A)[Y - a(A)] - \theta \Big)^2 \Big| A=v \Big\},
\eel
and hence also of \eqref{eq:bayes-risk-ab}, are
\bel{eq:oracle-a-b}
a^*_n(v) = \E(Y | A=v),\quad b^*_n(v) = \frac{v}{\Var(Y|A=v)}
\eel
and the minimum Bayes risk is 
\bel{eq:opt-risk}
R_n(\btheta,\hbtheta^{a^*_n,b^*_n}) = r_n(a^*_n, b^*_n) = \E\Big[A\big\{1-b^*_n(A)\big\}\Big].
\eel
Therefore, \eqref{eq:opt-risk} is a lower bound on the risk achievable by any estimator of the form \eqref{eq:lin-given-v}, and $\hbtheta^{a^*_n,b^*_n}$ is the 
optimal solution within the class. 
Note that any estimator of the form \eqref{eq:bayes-normal-normal} is also of the form \eqref{eq:lin-given-v}, hence the risk of the best (oracle) rule of the form \eqref{eq:lin-given-v} is no greater than the risk of the best rule of the form  \eqref{eq:bayes-normal-normal}. 
If $\xi$ and $A$ are independent, $a^*_n(v) = \E(Y | A=v) = \E(\xi | A=v) = \E(\xi), \quad b^*_n(v) = v/(v + \Var(\xi))$, and the oracles of the forms \eqref{eq:bayes-normal-normal} and \eqref{eq:lin-given-v} coincide.

Finally, we note that existing nonparametric empirical Bayes estimators, such as the semi-parametric estimator of \citet{xie2012sure} and the nonparametric method of \cite{jiang2010empirical}, target the best predictor $g(Y,A)$ of $\xi$ where $g$ is restricted to some nonparametric class of functions. 
While the optimal $g$ may indeed be a non-linear function of $Y$, these methods implicitly assume independence between $\xi$ and $A$, and might still suffer from the gap between the optimal predictor $g(Y,A)$ assuming independence, and the {true} Bayes rule, namely, $\E(\xi|Y,A)$. 
Therefore, in some cases the oracle rule of the form \eqref{eq:lin-given-v} might still have smaller risk than the oracle choice of $g$ computed assuming independence between $\xi$ and $A$. 

\section{Group-linear Shrinkage Methods} \label{sec:gl} 

Let $\bX, \btheta$ and $\bV$ be as in \eqref{eq:hetero-normal}. 
%
The estimator in the following lemma will serve as a building block for our group-linear estimator. 
Note in this estimator that $\bar{X}$ is used as an estimate of the overall group mean. 
In addition, the estimator is spherically symmetric as a function of $\bX - \bar{X}$. 
Similar estimators that center on a known mean, and variations, have been discussed in \citet[][Theorem 3]{brown1975estimation}, \citet{bock1975minimax}, \citet{berger1985statistical}, 
\citet[Theorem 5.7; although there are some typos]{lehmann1998theory}, \citet{tan2015improved} and elsewhere. 

\begin{lemma} \label{lem:spher-symm}
Let $\hbtheta^c$ be an estimator given by $\thetahat^c_i = X_i $ if $n=1$, and otherwise
\bel{eq:spher-symm}
\thetahat^c_i= X_i - \bhat(X_i-\Xbar),\quad \bhat = \min\big(1, c_n\Vbar/s_n^2\big)  \label{sse}
\eel
where $\Xbar = \sum_{i=1}^n X_i/n$, $\Vbar = \sum_{i=1}^n V_i/n$, 
$s_n^2 = \sum_{i=1}^n(X_i-\Xbar)^2/(n-1)$ and $c_n$ is a positive constant. 
Let $V_{\max}=\max_{i\le n}V_i$ and 
$
c_n^*\ =\   \{ [ (n - 3) - 2( V_{\max}/\Vbar - 1 ) ] /(n -1) \}_+ \ = \ \{1-2(V_{\max}/\Vbar)/(n-1)\}_+. 
$
Then for $0\le c_n\le 2c_n^*$, 
\begin{equation} \label{eq:sure}
{\small
\begin{aligned}
 \frac{1}{n}\sum_{i=1}^n \E\Big(\htheta_i - \theta_i\Big)^2 
\ \le \   \Vbar\left[ 1 - \left(1-1/n\right) \E\left\{(2c_n^*-c_n)\bhat
+ (2-2c_n^*+c_n-s_n^2/\Vbar)I_{\{ s_n^2/\Vbar \le c_n\}}\right\}\right]
\  \le \  \Vbar.
\end{aligned}
}
\end{equation}
\end{lemma}

\noindent {\it Remarks: }
\benu
\item The main reason for using $\Xbar$ is analytical simplicity. When $\theta_i$ are all equal, the MLE of 
the common mean is the weighted least squares estimate $(\sum_{i=1}^n X_i/V_i)/(\sum_{i=1}^n 1/V_i)$. 
\item In \eqref{eq:sure} note that when $s_n^2/\Vbar \ge c_n$, $(2c_n^*-c_n)\bhat = (2c_n^*-c_n)c_n\Vbar/s_n^2$ 
attains maximum at $c_n=c_n^*$. In the homoscedastic case $V_{\max} = \Vbar$ and $c_n^* = (n-3)/(n-1)$ is the usual constant for the James-Stein estimator that shrinks toward the sample mean. 
In the heteroscedastic case, for a version of the estimator above that shrinks toward zero, a sufficient condition for minimaxity appears in \citet{tan2015improved} as $0 \leq c_n \leq 2 \{ 1 - 2(V_{\max}/\Vbar)/n \}$. 
This is consistent with Lemma~\ref{lem:spher-symm}.
\item For one-way unbalanced ANOVA, $V_i=\sigma^2/n_i$ where $\sigma^2$ is the error variance and $n_i$ is the number of observations for the $i$-th sub-population. 
Suppose that $\sigma^2$ is unknown and that we have an unbiased estimator $\hsigma^2=S_k/k$ of $\sigma^2$ independent of the observations, where $S_k/\sigma^2\sim\chi^2_k$. 
Then replacing $V_i$ in the lemma with the corresponding estimates $\widehat{V}_i=\hsigma^2/n_i$, the same conclusion still holds with $0\le c_n(1+2/k)\le 2c_n^*$.



\eenu




We are now ready to introduce an empirical Bayes estimator, which employs the spherically symmetric estimator of Lemma \ref{lem:spher-symm} to mimic the oracle rule $\hbtheta^{a^*,b^*}$. 
When the number of distinct values $V_i$ is very small compared to $n$, a natural competitor of $\hbtheta^{a^*_n,b^*_n}$ is obtained by applying a James-Stein estimator separately to each group of homoscedastic observations. 
Under appropriate conditions, this estimator asymptotically approaches the oracle risk \eqref{eq:opt-risk}. 
The situation in the general heteroscedastic problem, when the number of distinct values $V_i$ is not very small compared to $n$, is not as obvious; still, the expression for the optimal function $a^*$ and $b^*$ in \eqref{eq:oracle-a-b} suggests grouping together observations with {\it similar} variances $V_i$, and then applying a spherically symmetric estimator separately to each group. 

Block-linear shrinkage has been suggested before for the homoscedastic case by \citet{cai1999adaptive} 
in the context of asymptotic adaptive wavelet estimation. 
However, the estimator of \citet{cai1999adaptive} is motivated from an entirely different perspective, and addresses a very different oracle rule (itself a blockwise rule) from the oracle associated with our procedure.  
See also \cite{ma2015adaptive}. 
For the heteroscedastic case, \citet{tansteinized} comments briefly that block shrinkage methods building on a minimax estimator can be considered to allow different shrinkage patterns for observations with different sampling variances; this is very much in line with our approach. 

\begin{definition}[Group-linear Empirical Bayes Estimator for a Heteroscedastic Mean] \label{def:group-linear}
Let $J_1,\ldots, J_m$ be disjoint intervals. 
For $k=1,...,m$ denote
\bes
\calI_k = \{i: V_i\in J_k\}, \ \ \ 
n_k= |\calI_k|,\ \ \ 
\Vbar_k =\sum_{i \in \calI_k}\frac{V_i}{n_k}, \ \ \ 
\Xbar_k = \sum_{i \in \calI_k}\frac{X_i}{n_k},\ \ \ 
s_k^2 =\sum_{i \in \calI_k}\frac{(X_i-\Xbar_k)^2}{n_k\vee 2-1}.
\ees
Define a corresponding group-linear estimator $\hbtheta^{GL}$ componentwise by
\bel{eq:group-linear}
\htheta^{GL}_i = 
\begin{cases}
X_i - \min\Big(1, c_k\Vbar_k/s_k^2 \Big)(X_i-\Xbar_k), & i\in \calI_k \\
X_i, & \text{otherwise}
\label{group-linear}
\end{cases}
\eel
and note that $\htheta_i=X_i$ when $V_i\not\in  \cup_{k=1}^m J_k$ or $V_i\in J_k$ for some $k$ with $c_k=0$. 
\end{definition}

\begin{theorem} \label{thm:group-linear}
For $\hbtheta = \hbtheta^{GL}$ in Definition~\ref{def:group-linear} with $c_k = \big\{1-2\big(\displaystyle \max_{i \in \calI_k}V_i/\Vbar_k\big)/(n_k-1)\big\}_+$ the following holds:
\begin{enumerate}
\item Under the Gaussian model \eqref{eq:hetero-normal} with deterministic $(\theta_i,V_i), i\le n$, the risk of $\hbtheta$ is no greater than that of the naive estimator $\bX$ and therefore $\hbtheta$ is minimax
\bel{eq:group-sure}
\frac{1}{n}\sum_{i=1}^n \E\Big(\htheta_i - \theta_i\Big)^2 
\le \frac{1}{n}\sum_{i=1}^n \E\Big(X_i - \theta_i\Big)^2  = \frac{1}{n}\sum_{i=1}^n V_i = \Vbar. 
\eel
\item 
Let $(X_i,\theta_i,V_i), i=1,\ldots,n$, be i.i.d. vectors from any fixed (with respect to $n$) population satisfying (\ref{eq:hetero-normal}). 
Let $(Y,\xi,A)$ be defined by \eqref{eq:joint-dist}; $r(a, b)$ as defined in \eqref{eq:bayes-risk-ab}; and $a^*$ and $b^*$ as defined in \eqref{eq:oracle-a-b}. 
Then
\bel{eq:asymp-opt-cond}
\frac{1}{n}\sum_{i=1}^n \E \Big[ \Big(\htheta_i-\theta_i\Big)^2 \Big| \bV \Big]
\le \frac{1}{n}\sum_{i=1}^n r(a^*,b^*|V_i) + o(1)
\label{thm1}
\eel
with $\bV = (V_1,...,V_n)$ and for any sequence $V_1, V_2, ...$ such that the following holds:
 
With $|J|$ being the length of interval $J$,
\bel{eq:local-cond}
\begin{gathered}
\max\limits_{1\leq k\leq m}|J_k|\rightarrow 0,\; \min\limits_{1\leq k\leq m} n_k\rightarrow\infty,\ \ \ \ \ \ \ \ 
a^*(v), b^*(v) \mbox{\,are\, uniformly\, continuous\,}\\
\limsup\limits_{n\rightarrow\infty} \frac{\sum_{i=1}^n V_i}{n}< \infty, \, \limsup\limits_{n\rightarrow\infty} \frac{\sum_{i=1}^n V_i I_{\{V_i\notin \cup_{k=1}^m J_k\}}}{n}=0\\
\label{condition}
\end{gathered}
\eel
\end{enumerate}
\end{theorem}
\noindent {\it Remarks on the second part of the theorem:} 
\benu
\item 
Note that when $(X_i,\theta_i,V_i)$ are i.i.d., then each triple is distributed as $(Y,\xi,A)$. 
We assumed that the `population' distribution $(Y,\xi,A)$ itself does not depend on $n$ (in which case $r(a,b)$ and $a^*,b^*$ indeed do not depend on $n$). 
A similar statement would still hold when the distribution of $(Y,\xi,A)$ depends on $n$, under the conditions that $\{a_n^*\}, \{ b_n^*\}$ are equicontinuous and $\{a_n^*\}$ is uniformly bounded for any given finite interval. 
Although not considered here, an analogue of the second part of the theorem could be stated for the nonrandom situation, $X_i|(\theta_i, V_i) \sim N(\theta_i, V_i), 1\leq i\leq n$ with deterministic $\theta_i$ and $V_i$. 
In this case, suppose that the empirical joint distribution ${G}_n$ of $\{ (\theta_i, V_i): 1\leq i \leq n \}$ has a limiting distribution $G$. 
Then if we define the risk for candidates $a_n, b_n$ to be computed with respect to $G$, our estimator enjoys $r(\ahat_n, \bhat_n) \to r(a^*, b^*)$ under appropriate conditions on $a^*, b^*$. 

\item 
The continuity of shrinkage factor and location $b^*(v), a^*(v)$ allows to borrow strength from neighboring observations with similar variances. 
To asymptotically mimic the performance of the oracle rule, $\max_{1\leq k\leq m}|J_k|\rightarrow 0,\; \min_{1\leq k\leq m} n_k\rightarrow\infty$ are necessary wherever shrinkage is needed. 
The only intrinsic assumption is $\limsup_{n\rightarrow\infty} \sum_{i=1}^n V_i/n< \infty$, essentially `equivalent' to bounded expectation of $A$. 
It ensures that $\max_{1\leq k\leq m}|J_k|\rightarrow 0,\; \min_{1\leq k\leq m} n_k\rightarrow\infty$ are satisfied when $\cup_{k=1}^m J_k$ are chosen to cover most of the observations, and at the same time $\limsup_{n\rightarrow\infty} \sum_{i=1}^n V_i I_{\{V_i\notin \cup_{k=1}^m J_k\}}/n=0$, which takes care of the remaining observations (large or isolated $V_i$), and guarantees that their contribution to the 
risk is negligible. 

\item 
A statement on Bayes risk, when expectation is taken over $\bV$ in \eqref{thm1}, can be obtained in a similar way by replacing the conditions on $\bV$ with bounded expectation of the random variable $A$. 
We skip this for simplicity. 
\eenu


\bigskip

For the i.i.d. situation of the second part of Theorem~\ref{thm:group-linear}, the case $r(a^*,b^*) = 0$ corresponds to $\xi = a^*(A)$, a deterministic function of $A$ (equivalently, $b^*(A) \equiv 1$). 
In this case the precision in estimating the function $a^*$
is crucial, 
and calls for a sharper result than \eqref{eq:asymp-opt-cond} regarding the rate of convergence of the excess risk. 
Noting that, trivially, $\xi = a^*(A)$ implies that $\E (\xi |A=v) = a^*(v)$,  
$
X_i | V_i \sim N(a^*(V_i),V_i)
$
is a nonparametric regression model, i.e., $\theta_i$ is a deterministic measurable function of $V_i$. 
In this case, the rate of convergence in (\ref{eq:asymp-opt-cond}) 
depends primarily on the smoothness of the function $a^*(v)$. 
In the homoscedastic case the smoothing feature of the James-Stein estimator was studied in \citet{li1984data}. 
The following theorem states that the group-linear estimator attains the optimal convergence rate under a Lipschitz condition, at least when $A$ is bounded.

\begin{theorem} \label{thm:group-linear-lip}
Let $(X_i,\theta_i,V_i), i=1,\ldots,n$, be i.i.d. vectors from a population satisfying (\ref{eq:hetero-normal}). 
If $r(a^*, b^*)=0$ and $a^*(\cdot)$ is $L$-Lipschitz continuous, then the group linear estimator in Definition~\ref{def:group-linear} with equal block size $|J_k|=|J|=\big (\frac{10V_{\max}^2}{nL}\big)^{\frac{1}{3}}$ and $c_n=c_n^*$ 
\bel{eq:asymp-opt-lip}
\frac{1}{n}\sum_{i=1}^n \E \Big[ \Big(\htheta_i-\theta_i\Big)^2 \Big| \bV \Big] \leq 2\Big(\frac{10V_{\max}^2\sqrt{L}}{n} \Big)^{\frac{2}{3}}
\eel
for any deterministic sequence $\bV = (V_1, ..., V_n)$. 
\end{theorem}

For the asymptotic results in Theorems~\ref{thm:group-linear} and~\ref{thm:group-linear-lip} to hold, it is enough to choose bins $J_k$ of equal length $|J| = \big (\frac{10V_{\max}^2}{nL}\big)^{\frac{1}{3}}$. 
However, in realistic situations, where $n$ is some fixed number, other strategies for binning observations according to the $V_i$ might be more sensible. 
For example, by Lemma~\ref{lem:spher-symm} and the first remark that follows it, bins that keep $ \big(\max \{V_i:i \in J_k\} \big) / \Vbar_k$ (rather than $\max \{V_i: i \in J_k\} - \min \{V_i:i \in J_k\}$) approximately fixed may be more appropriate. 
Hence we propose to bin observations to windows of equal lengths in $\log(V_i)$ instead of $V_i$. 
Furthermore, instead of the constant multiplying $n^{-1/3}$ in $|J|$, which may be suitable when the $V_i \in (0,1]$, we suggest in general to fix the {\it number} of bins to $n^{1/3}$, i.e., divide $\log(V_i)$ to bins of equal length $[\max_i (\log V_i) - \min_i (\log V_i)]/n^{1/3}$. 
On a finer scale, for a given choice of $\{ J_k \}$, there is also the question whether any two groups should be combined together, and the shrinkage factors adjusted accordingly; 
this issue arises even in the homoscedastic case \citep[][]{efron1973combining}. 
Note that, trivially, minimaxity is preserved when the values of $V_i$, but not $X_i$, are used to choose the bins $J_k$. 

As for performance of the group-linear estimator for fixed $n$, some situations are certainly harder than others. 
In the best scenario where the variances are clustered at a fixed finite set of possible values, the method is expected to work very well with fast convergence in \eqref{eq:asymp-opt-cond}. 
Otherwise, the method is expected to work reasonably well in the sense of \eqref{eq:asymp-opt-cond} when $\max V_i/\min V_i$ is not too large, whether the distribution of $V_i$ is continuous or not, because the large clusters will benefit from shrinkage and small clusters will have small total contribution to the risk due to minimaxity within each group. 
Still, the difference between the two cases could be nontrivial in finite samples. 
In the third and worst case scenario, the sequence of variances is rapidly increasing so that the benefit of grouping is small for a large fraction of relatively large variances. 
This could also happen when the variances are small, as the risk ratio between the group and naive estimators depends only on the ratio $V_i/V_{\max}$.

\section{Simulation Study} \label{sec:simulation} 
\begin{figure}[h]
\centering
 \includegraphics[scale=.8]{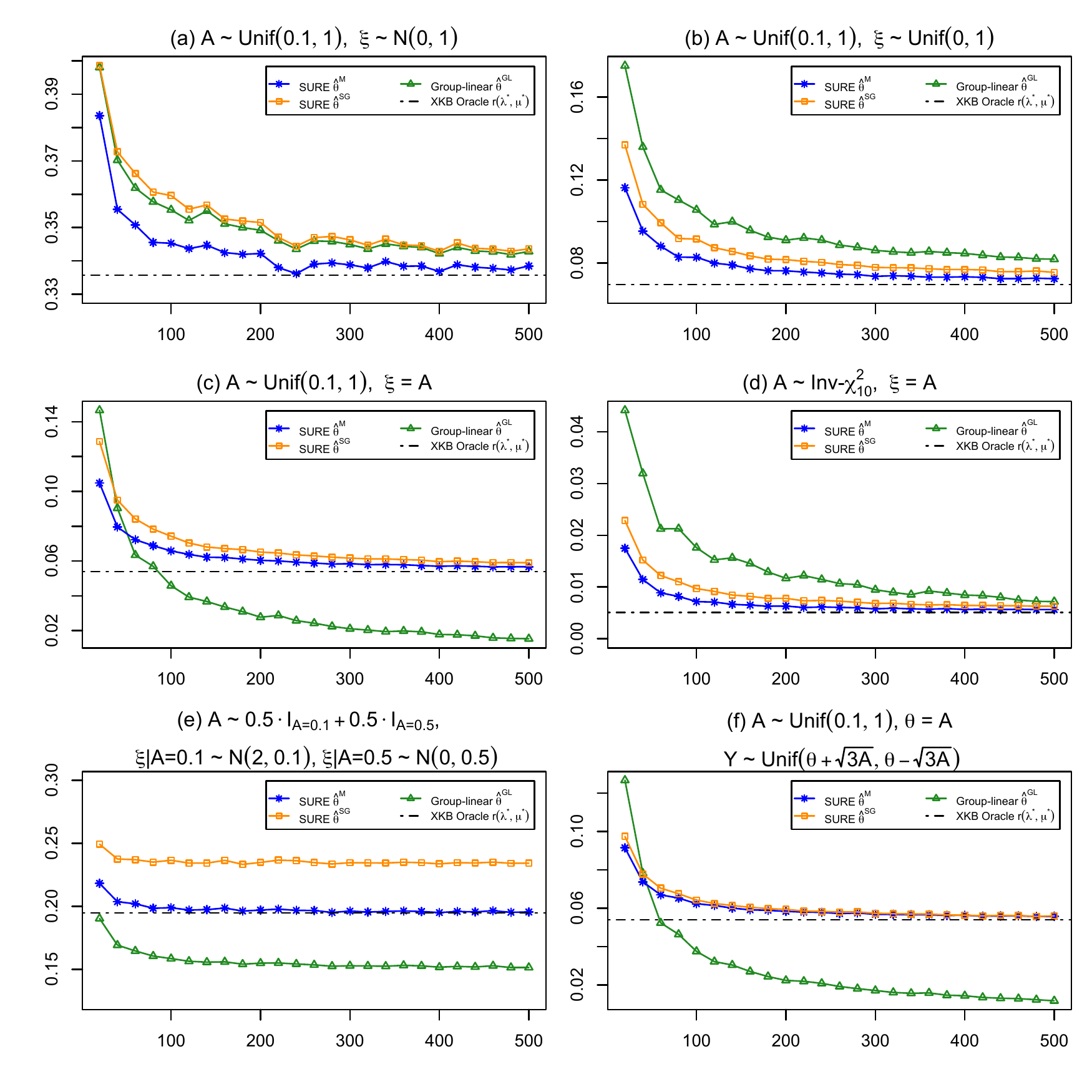}
\caption{ 
Estimated risk for various estimators vs. number of observations.
}
\label{fig:simulation}
\end{figure}

In this section we carry out a simulation study using the examples of \citet{xie2012sure}, and compare the performance of our group-linear estimator to the methods proposed in their work. 
In each example, we draw $n$ i.i.d. triples $(X_i, \theta_i, V_i) \sim (Y,\xi,A)$ such that $Y | (\xi, A)\sim N(\xi,A)$; 
the last example is the only exception, with $Y | (\xi, A)\nsim N(\xi,A)$, in order to assess the sensitivity to departures from normality. 
Various estimators are then applied to the data $(X_i, V_i), 1\leq i \leq n$, and the normalized sum of squared error is computed. 
For each value of $n$ in $\{ 20, 40, 60,...,500 \}$, this process is repeated $N=10,000$ times to obtain a good estimate of the (Bayes) risk for each method. 
Among the empirical Bayes estimators proposed by \citet{xie2012sure} we consider the parametric SURE estimator given by
\bes
\htheta_i^M = X_i - \frac{V_i}{V_i + \gammahat} (X_i - \hmu) , \ \ \ \ 1\leq i \leq n
\ees
where $\gammahat$ and $\hmu$ minimize an unbiased estimator of the risk (SURE) for estimators of the form 
$\htheta_i^{\mu, \gamma} = X_i - [V_i/(V_i + \gamma)] (X_i - \mu)$ over $\mu$ and $\gamma$. 
We also consider the semi-parametric SURE estimator of \citet{xie2012sure} with shrinkage towards the grand mean, defined by 
\bel{eq:sure-sg}
\htheta_i^{SG} = X_i - \bhat_i(X_i - \Xbar),\ \ \ \ 1\leq i \leq n
\eel
where $\hbb = (\bhat_1,...,\bhat_n)$ minimize an unbiased estimator of the risk for estimators of the form 
$\htheta_i^{\bb, \mu} = X_i - b_i(X_i - \Xbar)$ with $\bb = (b_1,...,b_n)$ restricted to satisfy $b_i \leq b_j$ whenever $V_i \leq V_j$. 
The group-linear estimator $\hbtheta^{GL}$ of Definition~\ref{def:group-linear} is applied here with the bins $J_k$ formed by dividing the range of $\log(V_i)$ into $\floor{n^{1/3}}$ equal length intervals, per the discussion concluding Section ~\ref{sec:gl}. 
As benchmarks, in each example we also compute the two oracle risks
\bel{eq:ol-xkb}
r(\mu^*, \gamma^*) = \displaystyle \min_{\mu, \gamma \in \R \ : \  \gamma\geq 0} \ \E \left\{ \big[   Y - \frac{A}{A + \gamma} (Y - \mu) - \xi   \big]^2 \right\}
\eel
and
\bel{eq:ol-lin-x}
r(a^*, b^*) = \displaystyle \min_{a(\cdot), b(\cdot) \ : \ a(v)\geq 0 \ \forall v} \ \E \left\{ \big[ Y - b(A) \big( Y - a(A) \big) - \xi \big]^2 \right\}
\eel
corresponding to the optimal rule in the parametric family of estimators considered in \citet[][labeled ``XKB oracle" in the legend of Figure ~\ref{fig:simulation}]{xie2012sure}, and to the optimal linear-in-$x$ rule of Section~\ref{sec:cd-hetero}, respectively.  
Note that $\mu^*$ and $\gamma^*$ are numbers whereas $a^*$ and $b^*$ are functions. 
Table \ref{table:simulation} displays the oracle shrinkage locations and shrinkage factors corresponding to \eqref{eq:ol-xkb} and \eqref{eq:ol-lin-x}; 
note that $v/(v + \gamma^*)$ is strictly increasing in $v$, while $b^*(v)$ is not necessarily. 

\begin{table}
  \begin{center}
\caption{Oracle shrinkage locations and shrinkage factors, $\mu^*, v/(v + \gamma^*)$ and $a^*(v), b^*(v)$, corresponding to the family of estimators of \citeauthor{xie2012sure} (equation 
(23)
) and to the family of estimators that are linear in $Y$ (equation 
(24) 
).
Columns correspond to simulation examples (a)- (f). 
Values of $\mu^*, \gamma^*$ for each example are from \citet{xie2012sure}. }  
\medskip
\begin{tabular}{ l l l l l l l } 
\hline \hline
  & (a) & (b) & (c) & (d) & (e) & (f) \\
  \hline 
  $\mu^*, \frac{v}{v + \gamma^*}$ & $0,\frac{v}{v+1}$ & $.5, \frac{v}{v+.083}$ & $0.6, \frac{v}{v + 0.078}$ & $0.13,\frac{v}{v + 0.0032}$ & $0.15, \frac{v}{v + 0.84}$ & $0.6, \frac{v}{v + 0.078}$ \\ 

  $a^*(v), \ b^*(v)$ & $0, \frac{v}{v + 1}$ & $0,\frac{v}{v + 1}$  & $v,0$ & $v,0$ & $ 2 \delta_{\{v=0.1\}}(v), 0.5 $ & $v,0$ \\
  \hline
\end{tabular}
\label{table:simulation}
  \end{center}
\end{table}

%

Figure~\ref{fig:simulation} shows the average loss across the $N=10,000$ repetitions for the parametric SURE, semi-parametric SURE and the group-linear estimators, plotted against the different values of $n$. 
The horizontal line corresponds to $r(\mu^*, \gamma^*)$. 
The general picture arising from the simulation examples is consistent with our expectation that the limiting risk of the group-linear estimator is smaller than that of both the parametric SURE estimator, as $r(a^*, b^*) \leq r(\mu^*, \gamma^*)$, and the semi-parametric SURE estimator, as $r(a^*, b^*) \leq \inf\{r(a,b): \text{$b(v)$ monotone increasing in $v$}\}$. 
For moderate $n$, whenever $\xi$ and $A$ are independent, the SURE estimators are appropriate and achieve smaller risk. 
By contrast, the situations where $\xi$ and $A$ are dependent are handled best by the group-linear estimator, which indeed achieves significantly smaller risk than both SURE estimators. 

%

In example (a) \citep[7.1 of][]{xie2012sure} $A\sim \text{Unif}(0.1,1)$ and $\xi \sim N(0,1)$, independently. 
In this case, the linear Bayes rule is of the form \eqref{eq:bayes-normal-normal} and, in particular, the functions $a^*$ and $b^*$ are constant in $v$. 
The parametric SURE estimator is therefore appropriate, and it performs best, requiring estimation of only two hyperparameters. 
The group-linear estimator and the semi-parametric SURE perform comparably across values of $n$. 
Here $r(\mu^*, \gamma^*), \ r(a^*, b^*)$ and the limiting risks of the parametric SURE and the group-linear estimator, are all equal ($\approx .3357$). 
In example (b) \citep[7.2 of][]{xie2012sure}, $A\sim \text{Unif}(0.1,1)$ and $\xi \sim N(0,1)$, independently. 
This situation is not very different from the first example when it comes to comparing the SURE estimators to the group-linear, since the functions $a^*$ and $b^*$ are constant in $v$ as long as $\xi$ and $A$ are independent. 
The risk of the group-linear approaches the oracle risk ($\approx .0697$), but here the semi-parametric SURE estimator seems to do a little better, perhaps in part because it (correctly) shrinks all data points toward exactly the same location. 

The third example (c) \citep[7.3 of][]{xie2012sure} takes $A\sim \text{Unif}(0.1,1)$, $\xi = A$. 
Here $\xi$ and $A$ are strongly dependent, and indeed the gap between the two oracle risks, $r(\mu^*, \gamma^*) \approx .0540$ and $r(a^*, b^*) = 0$, is material. 
The advantage of the group-linear estimator over the SURE estimators is seen already for moderate values of $n$. 
Although it is hard to tell from the figure, the limiting risk of the semi-parametric SURE is slightly smaller than that of the parametric SURE, because of the improved capability of the semi-parametric oracle to accommodate the dependence between $\xi$ and $A$. 
In the fourth case (d) \citep[7.3 of][]{xie2012sure} we take $A\sim \text{Inv-}\chi^2_{10}$, $\xi = A$. 
$\xi$ is still a deterministic function of $A$, but it takes larger values of $n$ for the group-linear estimator to outperform the SURE estimators. 
This is not seen before $n=500$, which seems to be a consequence of the non-uniform distribution of the $V_i$, and only partially mitigated by binning according to $\log(V_i)$. 

Example (e) \citep[7.5 of][]{xie2012sure} reflects grouping: $A$ equals $0.1$ or $0.5$ with equal probability; 
$\xi|(A=0.1) \sim N(2,0.1)$ and $\xi|(A=0.5) \sim N(0,0.5)$. 
In each of the two variance groups, 
the group-linear estimator reduces to a (positive-part) James-Stein estimator, and performs significantly better than the SURE estimators. 
While not plotted in the figure, the other semi-parametric SURE estimator of \citet{xie2012sure}, which uses a SURE criterion to choose also the shrinkage location, achieves significantly smaller risk than the SURE estimators considered here; 
still, its limiting risk is $16\%$ higher than that of the group-linear.  

Lastly, in (f) \citep[7.6 of][]{xie2012sure} $A\sim \text{Unif}(0.1,1)$, $\xi = A$ and $Y|A \sim \text{Unif}(\xi - \sqrt{3A}, \xi + \sqrt{3A})$, violating the normality assumption for the data. 
The group-linear estimator is again seen to outperform the SURE estimators starting at relatively small values of $n$, and its risk still tends to the oracle risk $r(a^*, b^*)=0$. 
By contrast, the risk of the parametric SURE estimator approaches $r(\mu^*, \gamma^*) = 0.054$. 
The semi-parametric SURE estimator does just a little better, its risk approaching $\approx 0.0423$. 

\section{Real Data Example} \label{sec:example}
We now turn to a real data example to test our group-linear methods. 
We use the popular baseball dataset from \cite{brown2008season}, which contains batting records for all Major League baseball players in the 2005 season. 
As in \cite{brown2008season}, the entire season is split into two periods, and the task is to predict the batting averages of individual players in the second half-season based on records from the first half-season. 
Denoting by $H_{ji}$ the number of hits and by $N_{ji}$ the number of at-bats for player $i$ in period $j$ of the season, it is assumed that
\bel{eq:bin}
H_{ji} \sim \text{Bin}(N_{ji}, p_i),\ \ \ \ \ j=1,2, \ i=1,...,\mathcal{P}_j.
\eel
As suggested in \cite{brown2008season}, a variance-stabilizing transformation is first applied,
$
X_{ji} = \arcsin \{ ( H_{ji} + 1/4 )^{1/2} / ( N_{ji} + 1/2 )^{1/2} \},
$
resulting in 
$
X_{ji} \stackrel{.}{\sim} N\big(\theta_i, 1/(4N_{ji}) \big),\ \theta_i = \arcsin(p_i), 
$
and $\{ (X_{1i}, N_{1i}): i=1,...,\calP_1 \}$ are then used to estimate the means $\theta_i$. 
We should remark that there is no reason for using this transformation, and for focusing on estimating $\theta_i$ instead of $p_i$, other than making the data (approximately) fit the heteroscedastic normal model (note that the variance of the obvious statistic $H_{ji} / N_{ji}$ depends explicitly on $p_i$, and therefore is not suitable). 
Indeed, one might object to analyzing the baseball data using a normal model  instead of using the binomial model \eqref{eq:bin} directly \citep[as in][]{muralidharan2010empirical}. 
Our only response is that the purpose of our analysis is primarily to illustrate the possible advantages of the group-linear estimator -- and more generally, of methods that can accommodate statistical dependence between the means and the known variances -- in the heteroscedastic normal problem. 

To measure the performance of an estimator $\hbtheta$, we use the Total Squared Error, 
$
\text{TSE}(\hbtheta) = \sum_{i} \left[ ( X_{2i} - \thetahat_i )^2 - 1/(4N_{2i}) \right], 
$
proposed by \cite{brown2008season} as an (approximately) unbiased estimator of the risk of $\hbtheta$. 
Following \cite{brown2008season}, only players with at least 11 at-bats in the first half-season are considered in the estimation process, and only players with at least 11 at-bats in both half-seasons are considered in the validation process, namely, when evaluating the TSE. 
To support our comparison, in addition to the analysis for the original data, we present an analysis under a permutation of the order in which successful hits appear throughout the entire season: for each player we draw the number of hits in the $N_{1i}$ at-bats of the first period from a Hypergeometric distribution, ${\cal H G}(N_{1i}+N_{2i},H_{1i}+H_{2i},N_{1i})$. 
In the permutation analysis we concentrate on the two SURE methods of \citet{xie2012sure}, which we consider as the main competitors of our method; the extended James-Stein estimator; and the group-linear estimators.

\begin{table} 
\centering
\caption{Prediction Errors of Transformed Batting Averages. 
For the five estimators at the bottom of the table, numbers in parentheses are estimated TSE for permuted data. } \bigskip
\resizebox{\columnwidth}{!}{
\begin{tabular}{ l c c c c c c  } 
\hline \hline
  & All & & Pitchers & & Non-pitchers & \\
  \hline
  Naive & 1 & & 1 &  & 1 &  \\
  Grand mean & .852 & & .127 & & .378 & \\
  Nonparametric EB & .508 &  & .212 &  & .372 &  \\
  Binomial mixture & .588 &  & .156 &  & .314 &  \\  
  Weighted Least Squares & 1.07 &  & .127 &  & .468 &  \\
  Weighted nonparametric MLE & .306 &  & .173 &  & .326 &  \\  
  Weighted Least Squares (AB) & .537 &  & .087 &  & .290 &  \\
  Weighted nonparametric MLE (AB) & {.301} &  & .141 &  & {\bf .261} &  \\  
  James-Stein & .535  & (.543) & .165 & (.239) & .348 & (.234) \\  
  SURE \ $\thetahat^M$ & .421 & (.484) & .123 & (.211)& .289 & (.265) \\    
  SURE \ $\thetahat^{SG}$ & .408 & (.468) & {\bf .091} & ({\bf .169}) & {\bf .261} & (.219) \\      
  Group-linear \ $\thetahat^{GL}$ & {.302} & (.280) & .178 & (.244) & .325 & ({\bf .175}) \\
  Group-linear (dynamic) & {\bf .288} & ({\bf .276}) & .168 & (.283) & .349 & ({\bf .175}) \\  
  \hline
\end{tabular}
}
\label{table:brown2008}
\end{table}

Table \ref{table:brown2008} shows TSE for various estimators reported in Table 2 of \citet{xie2012sure}, when applied separately to all players, pitchers only and non-pitchers only. 
The values displayed are fractions of the TSE of the naive estimator, which, in each of the cases (i)-(iii), simply predicts $X_{2i}$ by $X_{1i}$. 
Numbers in parentheses correspond to permuted data, and were computed as the average of the relative TSE over $1000$ rounds of shuffling as described above.  
In the table, the Grand mean estimator uses the simple average of all $X_{1i}$; 
the extended positive-part James-Stein estimator is given by 
$
\thetahat^{JS+}_i = \hat{\mu}_{JS+} + \big( 1-\frac{p-3}{\sum_i (X_i - \hat{\mu}_{JS+})} \big)_+(X_i - \hat{\mu}_{JS+})
$
where 
$\hat{\mu}_{JS+} = ({\sum_i X_i/V_i}) / ({\sum_i 1/V_i})$; 
$\hbtheta^M$ is the parametric empirical Bayes estimator of \citet{xie2012sure} using the SURE criterion to choose both the shrinkage and the location parameter; and 
$\hbtheta^{SG}$ is the semi-parametric SURE estimator of \citet{xie2012sure} that shrinks towards the grand mean. 
Also included in the table are the nonparametric shrinkage methods of \cite{brown2009nonparametric}; the weighted least squares estimator; the nonparametric maximum likelihood estimators of \cite{jiang2009general, jiang2010empirical} (with and without number of at-bats as covariate) and the binomial mixture estimator of \cite{muralidharan2010empirical}. 


For the group-linear estimator, in addition to the plain estimator $\hbtheta^{GL}$ that uses $k = \floor{n^{1/3}}$ equal length bins on $\log( \frac{1}{4N_{1i}} )$ (as in the simulation study), we considered a data-dependent strategy for binning.
The estimator labeled ``dynamic" in Table \ref{table:brown2008} chooses, among all partitions of the data into contiguous bins containing no more than $\floor{n^{2/3}}$ observations each, the partition which minimizes an unbiased estimate of the risk of the corresponding group-linear estimator. 
This can be viewed as an extension of the plain version, which for uniformly spaced data would put $\sim n^{2/3}$ observations in each of $\floor{n^{1/3}}$ bins. 
Our implementation uses dynamic programming (code available online at \url{https://github.com/MaZhuang/grouplinear}). 
We remark that using the observed data in forming the bins may lead to loss of minimaxity of the group-linear estimator. 
Nevertheless, we find it appropriate to explore such strategies when applying the estimator to real data.

Both versions of the group-linear estimator perform well in predicting batting averages for all players relative to the other estimators. 
As discussed in \cite{brown2008season}, nonconformity to the hierarchical normal-normal model on which most parametric empirical Bayes estimators are based, is evident in the data: 
first of all, non-pitchers tend to have better batting averages than pitchers, hence it is more plausible that the $\theta_i$ come from a {\it mixture} of two distributions. 
Second, players with higher batting averages tend to play more, suggesting that there is statistical dependence between the true means, $\theta_i$, and the sampling variances of $X_{ji}$ ($\propto 1/N_{ji}$); see Figure 4 in \citet{brown2008season}. 
While the nonparametric MLE method handles well non-normality in the ``prior" distribution of the $\theta_i$, its derivation still assumes statistical independence between the true means and the sampling variances. 
The group-linear estimator achieves good performance in this situation because it is able to accommodate this dependence between the true means and the sampling variances.

When analyzing pitchers and non-pitchers separately on the original data, the SURE methods achieve dramatic improvement, and outperform the group-linear estimators by a significant amount. 
However, the results are quite different for shuffled data. 
The difference is seen most prominently for non-pitchers: when actual second half records are used, the group-linear incurs higher prediction error as compared to the semi-parametric SURE estimator (0.325 vs. 0.261); but the opposite emerges for shuffled data (0.175 vs. 0.219). 
For pitchers only, the estimators of \citet{xie2012sure} outperform the group-linear in both the standard analysis and the permutation analysis. 
This is reasonable as the association between the number of at-bats and the true ability is expected to be weaker than within non-pitchers.

\section{Conclusion and Directions for Further Investigation} \label{sec:conclusion} 

For a heteroscedastic normal vector, empirical Bayes estimators that have been suggested, both parametric and nonparametric, usually rely on a hierarchical model in which the parameter $\theta_i$  has a prior distribution unrelated to the observed sampling variance $V_i = \Var(X_i|\theta_i)$. 
If separable estimators are considered, representing the heteroscedastic normal mean estimation problem as a compound decision problem, reveals that this model is generally inadequate to achieve risk reduction as compared to the naive estimator. 
Group-linear methods, on the other hand, are capable of capturing dependency between $\theta_i$ and $V_i$, and therefore are more appropriate for problems where it exists. 


There is certainly room for further research. 
We point out a few possible directions for extending Theorems~\ref{thm:group-linear} and~\ref{thm:group-linear-lip}, that are outside the scope of the current work:
\begin{enumerate}
\item In the i.i.d. case, the distribution of the population $(Y,\xi,A)$ may be allowed to depend on $n$ in such a way that $r_n(a^*_n, b^*_n) \to 0$ as $n\to \infty$. 
In this case the criterion (\ref{eq:asymp-opt-cond}) should be strengthened to 
the asymptotic ratio optimality criterion 
\bel{eq:ratio-opt}
\frac{1}{n}\sum_{i=1}^n \E\Big(\htheta_i-\theta_i\Big)^2
\le (1+o(1))r_n(a^*_n,b^*_n)
\eel
as $n\to\infty$. 
As (\ref{eq:ratio-opt}) does not hold uniformly for all joint distributions of $(Y,\xi,A)$, 
a reasonable target would be to prove \eqref{eq:ratio-opt} when $r_n(a^*,b^*)\ge \eta_n$ for small $\eta_n$ 
under suitable side conditions on the joint distribution of $(Y,\xi,A)$. 
This theory should include (\ref{eq:asymp-opt-cond}) as a special case 
and still maintain the property (\ref{eq:group-sure}).  

\item When $a^*(v)$ satisfies an order $\alpha$ smoothness condition with $\alpha>1$, 
a higher-order estimate of $a^*(V_i)$ needs to be used to achieve the optimal rate $n^{-\alpha/(2\alpha+1)}$ 
in the nonparametric regression case, $r(a^*,b^*)=0$, 
e.g., $\ahat(V_i)$ with an estimated polynomial $\ahat(v)$ for each $J_k$.  
We speculate that such a group-polynomial estimator might still always outperform the naive estimator 
$\htheta_i = X_i$ under a somewhat stronger minimum sample size requirement. 


\end{enumerate}


%
%
%

\appendix
\setcounter{secnumdepth}{0}

\section{Appendix: Proofs}
\newcommand{\remargin}{\vspace{-0.3cm}}
\paragraph{\textit{Proof of Lemma~\ref{lem:spher-symm}}}
It suffices to consider $0 < c_n\le 2c_n^*$. 
Let $b(x) = \min(1,c_n \Vbar/x)$ such that $\bhat = b\big(s_n^2\big)$. 
Notice that $(\pa/\pa X_i)s_n^2 = 2(X_i-\Xbar)/(n-1)$. By Stein's lemma,
\remargin
\bes
\E(X_i-\theta_i)(X_i-\Xbar)\bhat 
= V_i\,\E \Big\{(1-1/n)b(s_n^2) + 2(X_i-\Xbar)^2b'\big(s_n^2\big)/(n-1)\Big\}. 
\ees
\remargin
By definition, $2V_i/(n-1)\le \Vbar(1-c_n^*)$ and $xb'(x) = -b(x)I\{b(x)<1\}$, 
\bes
&& \frac{1}{n}\sum_{i=1}^n \E\Big( X_i - (X_i - \Xbar)\bhat - \theta_i\Big)^2 
\cr &=& \frac{1}{n}\sum_{i=1}^n\left[V_i + \E (X_i - \Xbar)^2b^2(s_n^2) 
-2V_i\E\left\{(1-1/n)b(s_n^2) + \frac{2(X_i-\Xbar)^2b'\big(s_n^2\big)}{n-1}\right\}\right]
\cr &\le & \Vbar + \left(1-1/n\right)\E \left\{s_n^2 b^2(s_n^2) - 2\Vbar b(s_n^2)
+\Vbar(1-c_n^*)2b\big(s_n^2\big)I_{\{ s_n^2 > c_n\Vbar\}}\right\}
\cr & = & \Vbar + \left(1-1/n\right) \E\, \Vbar b(s_n^2)\left\{ \min\big(s_n^2/\Vbar, c_n\big) - 2
+2(1-c_n^*) I_{\{ s_n^2 > c_n\Vbar\}}\right\}
\cr & = & \Vbar - \left(1-1/n\right) \E\, \Vbar b(s_n^2)\left\{ (2c_n^*-c_n) I_{\{ s_n^2 > c_n\Vbar\}}
+ (2-s_n^2/\Vbar)I_{\{ s_n^2 \le c_n\Vbar\}}\right\}
\cr & = & \Vbar\left[ 1 - \left(1-1/n\right) \E\left\{ b(s_n^2)(2c_n^*-c_n)
+ (2-2c_n^*+c_n-s_n^2/\Vbar)I_{\{ s_n^2/\Vbar \le c_n\}}\right\}\right]\leq \Vbar. 
\ees
Define $\epsilon_{|J|}=\max\limits_{v_1,v_2\in J} \big\{|a^*(v_1)-a^*(v_2)|, |b^*(v_1)-b^*(v_2)|\big\}, g(v)= \Var(\xi| A=v)$ and $h(v)=\mathbb{E}(\xi^2| A=v)$. 
Unless otherwise stated, all expectations and variances are conditional on $\bV$. 


\begin{lemma}[Analysis within each block]
Let $(X_i,\theta_i,V_i)_{i=1}^n$ be i.i.d. vectors drawn from some population $(Y,\xi,A)$ satisfying (\ref{eq:hetero-normal-cond}) with $n\geq 2$. 
If $\,V_1, \cdots, V_n\in J$ for some interval $J$ and $\min_{1\leq i\leq n}b^*(V_i)\geq \varepsilon, b^*(\Vbar)\geq \varepsilon\,$ for some $\varepsilon> 0$. Then the spherically symmetric shrinkage estimator defined in \eqref{group-linear} with $c_n=c_n^*$ satisfies
\begin{equation}
\begin{aligned}
\frac{1}{n}\sum_{i=1}^n \mathbb{E} \Big[ \Big(\htheta_i-\theta_i\Big)^2 \Big| \bV \Big] &\leq  \frac{1}{n}\sum_{i=1}^n r(a^*, b^*|V_i)+\frac{7V_{\max}}{n-1}+(\Vbar\epsilon_{|J|}+|J|)\frac{\varepsilon^2+1}{\varepsilon^2}+\epsilon^2_{|J|} \\&+\frac{2}{n-1}\Big\{\sum_{i=1}^n V_i^2+2\sum_{i=1}^n (V_i+\Vbar)h(V_i)+\Vbar^2\Big\}^{\frac{1}{2}}  
\end{aligned}
\end{equation}
where $V_{\max}=\max\{V_1, \cdots, V_n\}$ and $\Vbar=\sum_{i=1}^n V_i/n$.
\label{block}
\end{lemma}
\paragraph{\textit{Proof of Lemma~\ref{block}}}
As in the proof of Lemma 1 with $c_n=c_n^*$, 
\begin{align*}
\frac{1}{n}\sum_{i=1}^n \mathbb{E} \Big[ \Big(\htheta_i-\theta_i\Big)^2 \Big| \bV \Big]&=\frac{1}{n}\sum_{i=1}^n \mathbb{E}\Big(X_i-(X_i-\Xbar)\bhat-\theta_i  \Big| \bV \Big)^2\\
&\leq \Vbar + \left(1-\frac{1}{n}\right) \mathbb{E}\, \Vbar b(s_n^2)\left\{ \min\big(s_n^2/\Vbar , c_n^*\big) - 2+2(1-c_n^*) I_{\{ s_n^2 > c_n^*\Vbar\}}\right\}
\end{align*}
By definition, $r(a^*, b^*|V_i)=V_i(1-b^*(V_i))$ and $\min\big(s_n^2/\Vbar, c_n^*\big) \leq c_n^*\leq 1$. Then,
\begin{equation}
\frac{1}{n}\sum_{i=1}^n \mathbb{E} \Big[ \Big(\htheta_i-\theta_i\Big)^2 \Big| \bV \Big]\leq \frac{1}{n}\sum_{i=1}^n r(a^*, b^*|V_i)+\frac{1}{n}\sum_{i=1}^nb^*(V_i)V_i-\left(1-\frac{1}{n}\right)\Vbar\mathbb{E}(\bhat)+2\Vbar(1-c_n^*) \nonumber
\end{equation}
Observing that $0\leq \bhat\leq 1$ and $\Vbar (1-c_n^*)\leq 2V_{\max}/(n-1)$, 
\remargin
\begin{align*}
\frac{1}{n}\sum_{i=1}^n \mathbb{E}& \Big[ \Big(\htheta_i-\theta_i\Big)^2 \Big| \bV \Big]\leq \frac{1}{n}\sum_{i=1}^n r(a^*, b^*|V_i)+4V_{\max}/(n-1)+\Vbar/n+\frac{1}{n}\sum_{i=1}^nb^*(V_i)V_i-\Vbar\mathbb{E}(\bhat) \\
&\leq \frac{1}{n}\sum_{i=1}^n r(a^*, b^*|V_i) + 5V_{\max}/(n-1)+\Vbar \Big(\max\limits_{1\leq i\leq n}b^*(V_i)-\mathbb{E}\bhat \Big)\\
&= \frac{1}{n}\sum_{i=1}^n r(a^*, b^*|V_i) + 5V_{\max}/(n-1)+\Vbar\big\{\max\limits_{1\leq i\leq n}b^*(V_i)-b^*(\Vbar)\big\}+\Vbar  \Big(b^*(\Vbar)-\mathbb{E}\bhat \Big)\\
&\leq \frac{1}{n}\sum_{i=1}^n r(a^*, b^*|V_i) + 5V_{\max}/(n-1)+\Vbar\epsilon_{|J|}+\Vbar \Big(b^*(\Vbar)-\mathbb{E}\bhat \Big)
\end{align*}
where the last inequality is due to the uniformly continuity of $b^*(v)$. 
Next we will bound $\Vbar \Big(b^*(\Vbar)-\E\bhat \Big)$. By definition, $\Vbar(b^*(\Vbar)-\E\bhat)=\Vbar\mathbb{E}\Big\{ \Vbar/\Var(Y|A=\Vbar)-\min(1, c_n^*\Vbar/s_n^2)\Big\}$.
Further observe that $\Vbar/\Var(Y|A=\Vbar)=\Vbar/\left(\Vbar+\Var(\xi|A=\Vbar)\right)\leq 1$, 
\remargin
\begin{align*}
\Vbar(b^*(\Vbar)-E\bhat)&\leq \Vbar\mathbb{E}\Big\{ \Big(\Vbar/\Var(Y|A=\Vbar)-c_n^*\Vbar/s_n^2\Big)I_{\{c_n^*\Vbar\leq s_n^2\}}\Big\}\\
&\leq \Vbar\mathbb{E}\Big\{\left(1-c_n^*\Var(Y|A=\Vbar)/s_n^2\right)I_{\{c_n^*\Vbar\leq s_n^2\}}\Big\}\\
&=\mathbb{E} \Vbar\Big\{ (1-c_n^*)I_{\{c_n^*\Vbar\leq s_n^2\}}+\frac{c_n^*}{s_n^2}\big[s_n^2-\Var(Y|A=\Vbar)\big]I_{\{c_n^*\Vbar\leq s_n^2\}}\Big\}
\remargin
\end{align*}
Also, noting that $1-c_n^*\geq 0$ and $c_n^*\Vbar/s_n^2I_{\{c_n^*\Vbar\leq s_n^2\}}\leq 1$,
\remargin
\begin{align*}
\Vbar \Big(b^*(\Vbar)-E\bhat \Big)&\leq \Vbar(1-c_n^*)+\mathbb{E} |s_n^2-\Var(Y|A=\Vbar)|\\
&\leq 2V_{\max}/(n-1)+\mathbb{E} |s_n^2-\E s_n^2|+|\mathbb{E} s_n^2-\Var(Y|A=\Vbar)|\\
&= 2V_{\max}/(n-1)+\mathbb{E}\Big\{\mathbb{E}_{\btheta}|s_n^2-\E s_n^2| \Big\} +|\mathbb{E} s_n^2-\Var(Y|A=\Vbar)|\\
&\leq 2V_{\max}/(n-1)+\mathbb{E}\sqrt{\Var(s_n^2| \btheta)}+|\mathbb{E} s_n^2-\Var(Y|A=\Vbar)|\\
&\leq 2V_{\max}/(n-1)+ \Big\{\mathbb{E} \Big[\Var(s_n^2| \btheta)\Big]\Big\}^{\frac{1}{2}}+|\mathbb{E} s_n^2-\Var(Y|A=\Vbar)|
\remargin
\end{align*}
where the last two inequalities are due to Jensen's inequality. 
Conditionally on $\bV=(V_1, \cdots, V_n)$ and $\btheta=(\theta_1, \cdots, \theta_n)$, $\Xbar\sim N(\sum_{i=1}^n\theta_i/n, \sum_{i=1}^nV_i/n^2)$, and therefore
\vspace{-0.3cm}
\begin{align*}
\mathbb{E} (s_n^2)&=\frac{1}{n-1} \mathbb{E}\Big\{\mathbb{E} \Big(\sum_{i=1}^n X_i^2-n\Xbar^2|  \btheta \Big)\Big\} = \frac{1}{n-1}\mathbb{E}\Big\{\sum_{i=1}^n (V_i+\theta_i^2)-\frac{(\sum_{i=1}^n\theta_i)^2}{n}-\Vbar\Big\}\\
&=\Vbar+\frac{1}{n(n-1)}\Big\{(n-1)\sum_{i=1}^n \mathbb{E}(\xi^2|A=V_i)-\sum_{j\neq k} \mathbb{E}(\xi|A=V_j) \mathbb{E}(\xi| A=V_k)\Big\}
\tag{\stepcounter{equation}\theequation}\\
&=\Vbar+\frac{1}{n(n-1)}\Big\{(n-1)\sum_{i=1}^n \Var(\xi|A=V_i) + n\sum_{i=1}^n\big[\mathbb{E}(\xi|A=V_i)-\frac{1}{n}\sum_{j=1}^n\mathbb{E}(\xi|A=V_j)\big]^2 \Big\}\\
&\leq  \Vbar+\frac{1}{n}\sum_{i=1}^n \Var(\xi| A=V_i)+\frac{1}{n-1}\sum_{i=1}^n\big[\mathbb{E}(\xi|A=V_i)-\frac{1}{n}\sum_{j=1}^n\mathbb{E}(\xi|A=V_j)\big]^2\\
&=\Vbar+\frac{1}{n}\sum_{i=1}^n g(V_i)+\frac{1}{n-1}\sum_{i=1}^n\big[a^*(V_i)-\frac{1}{n}\sum_{j=1}^n a^*(V_j)\big]^2
\end{align*}
On the other hand, $\Var(Y|A=\Vbar)=\Vbar+\Var(\xi|A=\Vbar)=\Vbar+g(\Vbar)$. Hence,
\vspace{-0.3cm}
\begin{align*}
\vspace{-0.3cm}
|\mathbb{E}(s_n^2)-\Var(Y|A=\Vbar)|\leq \frac{1}{n}\sum_{i=1}^n|g(V_i)-g(\Vbar)|+\frac{1}{n-1}\sum_{i=1}^n\big[a^*(V_i)-\frac{1}{n}\sum_{j=1}^n a^*(V_j)\big]^2
\end{align*}
The uniform continuity of $a^*(v)$ implies that $|a^*(V_i)-\sum_{j=1}^n a^*(V_j)/n|\leq (n-1)/n\epsilon_{|J|}$. By definition, $b^*(v)=v/(v+g(v))$, then $g(v)=v/b^*(v)-v$ and therefore
\remargin
\begin{align*}
|g(V_i)-g(\Vbar)|&=\Big|\frac{V_ib^*(\Vbar)-\Vbar b^*(V_i)}{b^*(V_i)b^*(\Vbar)}+(V_i-\Vbar)\Big|\\
&\leq \frac{\big|V_i\big[b^*(\Vbar)-b^*(V_i)\big]\big|}{b^*(V_i)b^*(\Vbar)}+ \frac{\big| (V_i-\Vbar)b^*(V_i)\big|}{b^*(V_i)b^*(\Vbar)}+\big|V_i-\Vbar\big|
&\leq \frac{\left(V_i\epsilon_{|J|}+|J|\right)}{\varepsilon^2}+|J|
\end{align*}
where the last inequality follows from $\min_{1\leq i\leq n}b^*(V_i)\geq \varepsilon, b^*(\Vbar)\geq \varepsilon$. 
Combining the two inequalities above, $|\mathbb{E}(s_n^2)-\Var(Y|A=\Vbar)|\leq  \left(\Vbar\epsilon_{|J|}+|J|\right)/\varepsilon ^2+|J|+\epsilon^2_{|J|}$. Finally, we are going to control $\mathbb{E}\Big\{\Var(s_n^2| \btheta)\Big\}$. Again, $\Xbar| \bV, \btheta \sim N(\sum_{i=1}^n\theta_i/n, \sum_{i=1}^nV_i/n^2)$, hence
\remargin
\begin{align*}
\mathbb{E}\Big\{\Var(s_n^2| \btheta)\Big\}&=\frac{1}{(n-1)^2} \mathbb{E}\Big\{ \Var\Big(\sum_{i=1}^nX_i^2-n\Xbar^2| \btheta\Big)\Big\}\\
&\leq \frac{2}{(n-1)^2}\mathbb{E}\Big\{\Var\Big(\sum_{i=1}^nX_i^2|\btheta\Big)+\Var\Big(n\Xbar^2|\btheta\Big)\Big\}\\
&=\frac{2}{(n-1)^2}\mathbb{E}\Big\{\sum_{i=1}^n \left(2V_i^2+4\theta_i^2V_i\right)+n^2\left(2\Vbar^2/n^2+4\thetabar^2\Vbar/n\right) \Big\}
\end{align*}
By definition, $h(v)=\mathbb{E} (\xi^2|A=v)$, and, noting that $n\thetabar^2\leq \sum_{i=1}^n\theta_i^2$,
\remargin
\begin{equation}
\begin{aligned}
\mathbb{E}\Big\{\Var(s_n^2| \btheta)\Big\}&\leq \frac{4}{(n-1)^2}\Big\{\sum_{i=1}^n V_i^2+2\sum_{i=1}^n V_ih(V_i)+\Vbar^2+2\Vbar\sum_{i=1}^nh(V_i)\Big\} \\
&\leq \frac{4}{(n-1)^2}\Big\{\sum_{i=1}^n V_i^2+2\sum_{i=1}^n (V_i+\Vbar)h(V_i)+\Vbar^2\Big\}
\label{part2-lemma2}
\end{aligned}
\end{equation}
Put pieces together, we have
\begin{equation}
\begin{aligned}
\Vbar(b^*(\Vbar)-\mathbb{E}\bhat)\leq \frac{2V_{\max}}{n-1}+& \frac{\Vbar\epsilon_{|J|}+|J|}{\varepsilon^2}+|J|+\epsilon^2_{|J|} + \frac{2}{n-1}\Big\{\sum_{i=1}^n V_i^2+2\sum_{i=1}^n (V_i+\Vbar)h(V_i)+\Vbar^2\Big\}^{\frac{1}{2}} \nonumber
\end{aligned}
\end{equation}
\begin{equation}
\begin{aligned}
\frac{1}{n}\sum_{i=1}^n \mathbb{E} \Big[ \Big(\htheta_i-\theta_i\Big)^2 \Big| \bV \Big] &\leq \frac{1}{n}\sum_{i=1}^n r(a^*, b^*|V_i)+\frac{7V_{\max}}{n-1}+(\Vbar\epsilon_{|J|}+|J|)\frac{\varepsilon^2+1}{\varepsilon^2}+\epsilon^2_{|J|} \\
&+\frac{2}{n-1}\Big\{\sum_{i=1}^n V_i^2+2\sum_{i=1}^n (V_i+\Vbar)h(V_i)+\Vbar^2\Big\}^{\frac{1}{2}}  \nonumber
\end{aligned}
\end{equation}
\begin{proof}[\textbf{Proof of Theorem~\ref{thm:group-linear}}] 
The first part the Theorem follows from Lemma 1. 
For the second part, it suffices to show that for all $\varepsilon>0$, the excess risk is $O_n(\varepsilon)$. Notice that the contribution to the normalized risk for observations outside $\cup_{k=1}^m J_k$ is $\sum_{i=1}^n V_iI_{\{V_i\notin \cup_{k=1}^m J_k\}}/n=o(1)$, we only need to consider the case where $\forall 1\leq i\leq n, \, V_i\in \cup_{k=1}^m J_k$. Without loss of generality, we assume $\forall \, 1\leq k\leq m$, either $J_k\subset [0, \varepsilon)$ or $J_k\subset (\varepsilon, +\infty)$ since we can always reduce $\varepsilon$ such that this happens. Due to the assumption that $\limsup_{n\rightarrow\infty} \sum_{i=1}^n V_i/n< \infty$, we can also choose $M_\varepsilon$ large enough such that $\sum_{i=1}^n V_i I_{\{V_i\geq M_\varepsilon\}}/n\leq \varepsilon$ and for any $k$ with $J_k\subset  (\varepsilon, +\infty)$, either $J_k\subset  (\varepsilon, M_\varepsilon)$ or $J_k\subset (M_\varepsilon, +\infty)$.

For the rest of the proof, we divide all the observations into four disjoint groups and handle them separately. Let $\Vbar^k=\sum_{i\in \mathcal{I}_k }V_i/n_k$ and define $S_1=\{k| 1\leq k\leq n, J_k\subset (0, \varepsilon)\}, S_2=\{k| 1\leq k\leq n, J_k\subset (\varepsilon, M_\varepsilon), \min_{V_i\in J_k}b^*(V_i)\geq \varepsilon, b^*(\Vbar^k)\geq \varepsilon\}, S_3=\{k| 1\leq k\leq n, J_k\subset (\varepsilon, M_\varepsilon), \min_{V_i\in J_k}b^*(V_i)< \varepsilon\; or\;  b^*(\Vbar^k)\leq \varepsilon\}, S_4=\{k| 1\leq k\leq n, J_k\subset (M_\varepsilon, +\infty)\}$. 
 {\bf Case i)} For the small variance part, $V_i\in  (0, \varepsilon)$, the contribution to the risk is negligible. Because the group linear shrinkage estimator dominate the MLE in each interval, then
\begin{align*}
\frac{1}{n}\sum_{k\in S_1}\sum_{i\in\mathcal{I}_k}\mathbb{E} \Big[ \Big(\htheta_i-\theta_i\Big)^2 \Big| \bV \Big]\leq \sum_{k\in S_1}\sum_{i\in\mathcal{I}_k} V_i/n\leq \sum_{k\in S_1}\sum_{i\in\mathcal{I}_k}\varepsilon/n\leq \varepsilon
\end{align*}
{\bf Case ii)} For moderate variance with large shrinkage factor, $V_i\in  (\varepsilon, M_\varepsilon)$ and $b^*(V_i), b^*(\Vbar)\geq \varepsilon$, shrinkage is necessary to mimic the oracle.  
Applying Lemma~\ref{block} to each interval $J_k, k\in S_2$, 
\begin{align*}
\frac{1}{n}&\sum_{k\in S_2}\sum_{i\in\mathcal{I}_k}\mathbb{E} \Big[ \Big(\htheta_i-\theta_i\Big)^2 \Big| \bV \Big]\leq \frac{1}{n}\sum_{k\in S_2} \sum_{i\in\mathcal{I}_k} r(a^*, b^*|V_i)+\frac{1}{n}\sum_{k\in S_2} n_k\Big\{\frac{7}{n_k - 1}(\Vbar^k+|J_k|)\\&+\left(\Vbar^k\epsilon_{|J_k|}+|J_k|\right)\frac{\varepsilon^2+1}{\varepsilon^2}+\epsilon^2_{|J_k|}+\frac{2}{n_k - 1}\Big(\sum_{i\in\mathcal{I}_k} V_i^2+2\sum_{i\in\mathcal{I}_k} (V_i+\Vbar^k)h(V_i)+(\Vbar^k)^2\Big)^{\frac{1}{2}}\Big\}
\end{align*}
Let $|J|_{\max}=\max\limits_{1\leq k\leq m} |J_k|, \epsilon_{\max}=\max\limits_{1\leq k\leq m} \epsilon_{|J_k|}$. 
Using the fact that $\max\limits_{1\leq k\leq m} n_k/(n_k - 1)\leq 2$,
\begin{align*}
\frac{1}{n}&\sum_{k\in S_2}\sum_{i\in\mathcal{I}_k}\mathbb{E} \Big[ \Big(\htheta_i-\theta_i\Big)^2 \Big| \bV \Big]\leq \frac{1}{n}\sum_{k\in S_2} \sum_{i\in\mathcal{I}_k} r(a^*, b^*|V_i)+\frac{1}{n}\sum_{k\in S_2}\Big\{14(\Vbar^k+ |J|_{\max})+n_k\epsilon_{\max}^2\\
&+n_k(\Vbar^k\epsilon_{\max}+|J|_{\max})\frac{\varepsilon^2+1}{\varepsilon^2}+4\Big(\sum_{i\in\mathcal{I}_k} V_i^2+2\sum_{i\in\mathcal{I}_k} (V_i+\Vbar^k)h(V_i)+(\Vbar^k)^2\Big)^{\frac{1}{2}}\Big\}
\end{align*}
For any $k\in S_2, i\in \mathcal{I}_k, \, \Vbar^k, V_i\leq M_\varepsilon$. 
Because $a^*(v)$ is uniformly continuous on $[0, M_\varepsilon]$, there exists constant $C_\varepsilon$ only depending on $\varepsilon$ such that $a^*(V_i)\leq C_\varepsilon$. Then,
\begin{equation}
\begin{gathered}
	h(V_i)=\Var(\xi|A=V_i)+ \Big(\mathbb{E}(\xi|A=V_i) \Big)^2\leq V_i/b^*(V_i)-V_i+(a^*(V_i))^2\leq M_\varepsilon/\varepsilon+C_\varepsilon^2
\end{gathered}
\nonumber
\end{equation} 
\begin{equation}
		\begin{aligned}
\frac{1}{n}&\sum_{k\in S_2}\sum_{i\in\mathcal{I}_k}\mathbb{E} \Big[ \Big(\htheta_i-\theta_i\Big)^2 \Big| \bV \Big]\leq \frac{1}{n}\sum_{k\in S_2} \sum_{i\in\mathcal{I}_k} r(a^*, b^*|V_i)+\frac{14|S_2|}{n}\Big(M_\varepsilon+|J|_{\max} \Big)+\epsilon_{\max}^2\\
&+(M_\varepsilon\epsilon_{\max}+|J|_{\max})\frac{\varepsilon^2+1}{\varepsilon^2}+\frac{4}{n}\sqrt{2M_\varepsilon^2(1+\varepsilon^{-1})+2M_\varepsilon C_\varepsilon}\sum_{k\in S_2} \sqrt{n_k} 		
	\end{aligned}
	\nonumber
\end{equation}
By the Cauchy Schwarz inequality: $ \sum_{k\in S_2} \sqrt{n_k} \leq \sqrt{|S_2|\sum_{k\in S_2}n_k}\leq \sqrt{|S_2|n}$. Further observe that $|S_2|\leq m\leq n/\min\limits_{1\leq k\leq m} n_k$, then
\remargin
\begin{align*}
\frac{1}{n}&\sum_{k\in S_2}\sum_{i\in\mathcal{I}_k}\mathbb{E} \Big[ \Big(\htheta_i-\theta_i\Big)^2 \Big| \bV \Big]\leq \frac{1}{n}\sum_{k\in S_2} \sum_{i\in\mathcal{I}_k} r(a^*, b^*|V_i)+\frac{14}{\min\limits_{1\leq k\leq m}n_k}\Big(M_\varepsilon+|J|_{\max} \Big)+\epsilon_{\max}^2\\
&+(M_\varepsilon\epsilon_{\max}+|J|_{\max})\frac{\varepsilon^2+1}{\varepsilon^2}+\frac{4}{\sqrt{\min\limits_{1\leq k\leq m}n_k}}\sqrt{2M_\varepsilon^2(1+\varepsilon^{-1})+2M_\varepsilon C_\varepsilon}
\vspace{-0.3cm}
\end{align*}
Since $|J|_{\max}, \epsilon_{\max}\rightarrow 0$ and $\min\limits_{1\leq k\leq m}n_k\rightarrow +\infty$, we obtain \vspace{-0.3cm}
\begin{equation}
\frac{1}{n}\sum_{k\in S_2}\sum_{i\in\mathcal{I}_k}\mathbb{E} \Big[ \Big(\htheta_i-\theta_i\Big)^2 \Big| \bV \Big]\leq \frac{1}{n}\sum_{k\in S_2} \sum_{i\in\mathcal{I}_k} r(a^*, b^*|V_i)+o(\varepsilon) \nonumber
\vspace{-0.3cm}
\end{equation}
{\bf Case iii)} For moderate variance with negligible shrinkage factor, $V_i\in (\varepsilon, M_\varepsilon)$ and $\min_{i\in\mathcal{I}_k} b^*(V_i)\; or\; b^*(\Vbar)<\varepsilon$. The uniform continuity of $b^*(\cdot)$ implies that $\forall i\in\mathcal{I}_k,\; b^*(V_i)\leq \varepsilon+\epsilon_{\max}$. By definition $r(a^*, b^*|V_i)=V_i(1-b^*(V_i))$, then
\remargin
\begin{equation}
\frac{1}{n}\sum_{k\in S_3} \sum_{i\in\mathcal{I}_k} r(a^*, b^*|V_i)=\frac{1}{n}\sum_{k\in S_3} \sum_{i\in\mathcal{I}_k} V_i(1-b^*(V_i))\geq \frac{1}{n}\sum_{k\in S_3} \sum_{i\in\mathcal{I}_k} V_i-\Vbar(\varepsilon+\epsilon_{\max})\nonumber
\end{equation}
Since the proposed group linear shrinkage estimator dominates MLE in each block,
\begin{equation}
\begin{aligned}
\frac{1}{n}\sum_{k\in S_3}\sum_{i\in\mathcal{I}_k}\mathbb{E} \Big[ \Big(\htheta_i-\theta_i\Big)^2 \Big| \bV \Big]\leq \frac{1}{n}\sum_{k\in S_3} \sum_{i\in\mathcal{I}_k} r(a^*, b^*|V_i)+\Vbar(\varepsilon+\epsilon_{\max})\nonumber
\end{aligned}
\remargin
\end{equation}
{\bf Case iv)} For the large variance part, $V_i\in  (M_\varepsilon, +\infty)$, by the definition of $M_\varepsilon$,
\remargin
\begin{align*}
\frac{1}{n}\sum_{k\in S_4}\sum_{i\in\mathcal{I}_k}\mathbb{E} \Big[ \Big(\htheta_i-\theta_i\Big)^2 \Big| \bV \Big]\leq \sum_{k\in S_4}\sum_{i\in\mathcal{I}_k} V_i/n=\sum_{i=1}^n V_i I_{\{V_i\geq M_\varepsilon\}}/n\leq \varepsilon
\remargin
\end{align*}
Summing up the inequalities of all four cases
\remargin
$$\frac{1}{n}\sum_{i=1}^n\mathbb{E} \Big[ \Big(\htheta_i-\theta_i\Big)^2 \Big| \bV \Big]\leq \frac{1}{n}\sum_{i=1}^n r(a^*, b^*|V_i)+(\Vbar+2)\varepsilon+o(\varepsilon)$$
which completes the proof by the assumption that $\limsup\limits_{n\rightarrow \infty}\sum_{i=1}^n V_i/n\leq \infty$
\end{proof}

\begin{lemma}[Analysis within each block]
\label{block2}
Let $(X_i,\theta_i,V_i)_{i=1}^n$ be i.i.d. vectors from  some population $(Y,\xi,A)$ satisfying (\ref{eq:hetero-normal-cond}). 
If $r(a^*, b^*)=0, a^*(\cdot)$ is $L$-Lipschitz continuous and $V_1, \cdots, V_n\in J$ for some interval $J$, then the estimator defined in \eqref{group-linear}  with $c_n=c_n^*$ satisfies
\begin{equation}
\frac{1}{n}\sum_{i=1}^n \mathbb{E} \Big[ \Big(\htheta_i-\theta_i\Big)^2 \Big| \bV \Big]\leq L|J|^2+3\Vbar/n+4V_{\max}/(n\vee 2-1) \nonumber
\end{equation}
\end{lemma}
\paragraph{\textit{Proof of Lemma~\ref{block2}}}
As in the proof of Lemma~\ref{lem:spher-symm} and substitute $c_n$ with $c_n^*$
\remargin
\begin{align*}
& ~~~\frac{1}{n}\sum_{i=1}^n \mathbb{E} \Big[ \Big(\htheta_i-\theta_i\Big)^2 \Big| \bV \Big]=\frac{1}{n}\sum_{i=1}^n \mathbb{E}\left(X_i-(X_i-\Xbar)\bhat-\theta_i| \bV\right)^2\\
&\leq  \Vbar\left[ 1 - \left(1-1/n\right) \mathbb{E}\left\{ \bhat(2c_n^*-c_n)
+ (2-2c_n^*+c_n-s_n^2/\Vbar)I_{\{ s_n^2/\Vbar \le c_n\}}\right\}\right]\\
&=\Vbar\left[ 1 - \left(1-1/n	\right) \mathbb{E}\left\{ \bhat c_n^*
+ (2-c_n^*-s_n^2/\Vbar)I_{\{ s_n^2/\Vbar \le c_n^*\}}\right\}\right]\\
&=\Vbar \mathbb{E}\left\{ (1-\bhat c_n^*)
-(2-2c_n^*)I_{\{ s_n^2/\Vbar \le c_n^*\}}-(c_n^*-s_n^2/\Vbar)I_{\{ s_n^2/\Vbar \le c_n^*\}}\right\}\\
&\quad\quad\quad\quad\quad\quad+ \mathbb{E}\left\{ \bhat c_n^*+ (2-c_n^*-s_n^2/\Vbar)I_{\{ s_n^2/\Vbar \le c_n^*\}}\right\}\Vbar/n
\end{align*}
\remargin
Notice that $2-2c_n^*>0$ and $\bhat c_n^* + (2-c_n^*-s_n^2/\Vbar)I_{\{ s_n^2/\Vbar \le c_n^*\}}\leq 2$.
\remargin 
\begin{align*}
& ~~~\frac{1}{n}\sum_{i=1}^n \mathbb{E} \Big[ \Big(\htheta_i-\theta_i\Big)^2 \Big| \bV \Big]\leq \Vbar\mathbb{E}\left\{ (1-\bhat c_n^*)
-(c_n^*-s_n^2/\Vbar)I_{\{ s_n^2/\Vbar \le c_n^*\}}\right\}+2\Vbar/n\\
&\leq \Vbar\mathbb{E}\Big\{c_n^*(1-\bhat)-(c_n^*-s_n^2/\Vbar)I_{\{ s_n^2/\Vbar \le c_n^*\}}\Big\}+2\Vbar/n+(1-c_n^*)\Vbar\\
&\leq \mathbb{E}\Big\{c_n^*\Vbar\left(\frac{s_n^2-c_n^*\Vbar}{s_n^2}\right)_+-\big(c_n^*\Vbar-s_n^2\big)_+\Big\}+2\Vbar/n+(1-c_n^*)\Vbar\\
&\leq \mathbb{E}\Big\{\big(s_n^2-c_n^*\Vbar\big)_+-\big(c_n^*\Vbar-s_n^2\big)_+\Big\}+2\Vbar/n+(1-c_n^*)\Vbar\\
&=\mathbb{E}(s_n^2-c_n^*\Vbar)+2\Vbar/n+(1-c_n^*)\Vbar
\end{align*}
Recall that {\small $\mathbb{E} s_n^2= \Vbar+\frac{1}{n}\sum_{i=1}^n \Var(\xi| A=V_i)+\frac{1}{n\vee 2-1}\sum_{i=1}^n[\E(\xi|A=V_i)-\frac{1}{n}\sum_{j=1}^n\E(\xi|A=V_j)\big]^2 \nonumber$}. With $\Var(\xi| A=v)=0$, we have $\E s_n^2= \Vbar+\frac{1}{n\vee 2-1}\sum_{i=1}^n[a(V_i)-\frac{1}{n}\sum_{j=1}^na(V_j)\big]^2$ and
\remargin
\begin{align*}
\frac{1}{n}\sum_{i=1}^n \mathbb{E} \Big[ \Big(\htheta_i-\theta_i\Big)^2 \Big| \bV \Big]&\leq 2(1-c_n^*)\Vbar+\frac{1}{n\vee 2-1}\sum_{i=1}^n[a(V_i)-\frac{1}{n}\sum_{j=1}^na(V_j)\big]^2+2\Vbar/n\\
&\leq L|J|^2+2\Vbar/n+2(1-c_n^*)\Vbar\leq L|J|^2+2\Vbar/n+\frac{4V_{\max}}{n\vee 2-1}
\end{align*}
\begin{proof}[\textbf{Proof of Theorem 2.}]
Apply Lemma~\ref{block2} to each interval and notice $n_k/(n_k - 1)\leq 2$, 
\remargin
\begin{align*}
\frac{1}{n}\sum_{i=1}^n \mathbb{E} \Big[ \Big(\htheta_i-\theta_i\Big)^2 \Big| \bV \Big]&\leq \frac{1}{n}\sum_{k=1}^m \big(n_kL|J_k|^2+2\Vbar^k+4V_{\max}\frac{n_k}{n_k\vee 2-1}\big)\\
&\leq L|J|^2+10mV_{\max}/n=L|J|^2+10V_{\max}^2/(n|J|)
\end{align*}
Letting $|J|=\big (\frac{10V_{\max}^2}{nL}\big)^{\frac{1}{3}}$, we have that
$\frac{1}{n}\sum_{i=1}^n\mathbb{E} \Big[ \Big(\htheta_i-\theta_i\Big)^2 \Big| \bV \Big]\leq 2\big(\frac{10V_{\max}^2\sqrt{L}}{n}\big)^{\frac{2}{3}}$
\end{proof}

\bibliography{References.bib}

\begin{thebibliography}{20}
\providecommand{\natexlab}[1]{#1}
\providecommand{\url}[1]{\texttt{#1}}
\expandafter\ifx\csname urlstyle\endcsname\relax
  \providecommand{\doi}[1]{doi: #1}\else
  \providecommand{\doi}{doi: \begingroup \urlstyle{rm}\Url}\fi

\bibitem[Berger(1976)]{berger1976admissible}
James~O Berger.
\newblock Admissible minimax estimation of a multivariate normal mean with
  arbitrary quadratic loss.
\newblock \emph{The Annals of Statistics}, pages 223--226, 1976.

\bibitem[Berger(1982)]{berger1982selecting}
James~O Berger.
\newblock Selecting a minimax estimator of a multivariate normal mean.
\newblock \emph{The Annals of Statistics}, 10\penalty0 (1):\penalty0 81--92,
  1982.

\bibitem[Berger(1985)]{berger1985statistical}
James~O Berger.
\newblock \emph{Statistical decision theory and Bayesian analysis}.
\newblock Springer, 1985.

\bibitem[Bock(1975)]{bock1975minimax}
Mary~Ellen Bock.
\newblock Minimax estimators of the mean of a multivariate normal distribution.
\newblock \emph{The Annals of Statistics}, pages 209--218, 1975.

\bibitem[Brown(1975)]{brown1975estimation}
Lawrence~D Brown.
\newblock Estimation with incompletely specified loss functions (the case of
  several location parameters).
\newblock \emph{Journal of the American Statistical Association}, 70\penalty0
  (350):\penalty0 417--427, 1975.

\bibitem[Brown(2008)]{brown2008season}
Lawrence~D Brown.
\newblock In-season prediction of batting averages: A field test of empirical
  bayes and bayes methodologies.
\newblock \emph{The Annals of Applied Statistics}, pages 113--152, 2008.

\bibitem[Brown and Greenshtein(2009)]{brown2009nonparametric}
Lawrence~D Brown and Eitan Greenshtein.
\newblock Nonparametric empirical bayes and compound decision approaches to
  estimation of a high-dimensional vector of normal means.
\newblock \emph{The Annals of Statistics}, pages 1685--1704, 2009.

\bibitem[Cai(1999)]{cai1999adaptive}
T~Tony Cai.
\newblock Adaptive wavelet estimation: a block thresholding and oracle
  inequality approach.
\newblock \emph{Annals of statistics}, pages 898--924, 1999.

\bibitem[Efron and Morris(1973{\natexlab{a}})]{efron1973combining}
Bradley Efron and Carl Morris.
\newblock Combining possibly related estimation problems.
\newblock \emph{Journal of the Royal Statistical Society. Series B
  (Methodological)}, pages 379--421, 1973{\natexlab{a}}.

\bibitem[Efron and Morris(1973{\natexlab{b}})]{efron1973stein}
Bradley Efron and Carl Morris.
\newblock Stein's estimation rule and its competitorsÑan empirical bayes
  approach.
\newblock \emph{Journal of the American Statistical Association}, 68\penalty0
  (341):\penalty0 117--130, 1973{\natexlab{b}}.

\bibitem[Jiang and Zhang(2009)]{jiang2009general}
Wenhua Jiang and Cun-Hui Zhang.
\newblock General maximum likelihood empirical bayes estimation of normal
  means.
\newblock \emph{The Annals of Statistics}, 37\penalty0 (4):\penalty0
  1647--1684, 2009.

\bibitem[Jiang and Zhang(2010)]{jiang2010empirical}
Wenhua Jiang and Cun-Hui Zhang.
\newblock Empirical bayes in-season prediction of baseball batting averages.
\newblock In \emph{Borrowing Strength: Theory Powering Applications--A
  Festschrift for Lawrence D. Brown}, pages 263--273. Institute of Mathematical
  Statistics, 2010.

\bibitem[Johnstone(2011)]{johnstone2011gaussian}
Iain~M Johnstone.
\newblock Gaussian estimation: Sequence and wavelet models.
\newblock \emph{Unpublished manuscript}, 2011.

\bibitem[Lehmann and Casella(1998)]{lehmann1998theory}
Erich~Leo Lehmann and George Casella.
\newblock \emph{Theory of point estimation}, volume~31.
\newblock Springer, 1998.

\bibitem[Li and Hwang(1984)]{li1984data}
Ker-Chau Li and Jiunn~Tzon Hwang.
\newblock The data-smoothing aspect of stein estimates.
\newblock \emph{The Annals of Statistics}, pages 887--897, 1984.

\bibitem[Ma et~al.(2015)Ma, Foster, and Stine]{ma2015adaptive}
Zhuang Ma, Dean Foster, and Robert Stine.
\newblock Adaptive monotone shrinkage for regression.
\newblock \emph{arXiv preprint arXiv:1505.01743}, 2015.

\bibitem[Muralidharan(2010)]{muralidharan2010empirical}
Omkar Muralidharan.
\newblock An empirical bayes mixture method for effect size and false discovery
  rate estimation.
\newblock \emph{The Annals of Applied Statistics}, 4\penalty0 (1):\penalty0
  422--438, 2010.

\bibitem[Tan(2014)]{tansteinized}
Zhiqiang Tan.
\newblock Steinized empirical bayes estimation for heteroscedastic data.
\newblock \emph{Preprint}, 2014.

\bibitem[Tan(2015)]{tan2015improved}
Zhiqiang Tan.
\newblock Improved minimax estimation of a multivariate normal mean under
  heteroscedasticity.
\newblock \emph{Bernoulli}, 21\penalty0 (1):\penalty0 574--603, 02 2015.

\bibitem[Xie et~al.(2012)Xie, Kou, and Brown]{xie2012sure}
Xianchao Xie, SC~Kou, and Lawrence~D Brown.
\newblock Sure estimates for a heteroscedastic hierarchical model.
\newblock \emph{Journal of the American Statistical Association}, 107\penalty0
  (500):\penalty0 1465--1479, 2012.

\end{thebibliography}
\bibliographystyle{plain}

\end{document}